\documentclass[twocolumn]{aastex631}

\received{XXX}
\revised{YYY}
\accepted{ZZZ}

\shorttitle{Secondary jet in OJ287}
\shortauthors{Valtonen et al.}

\graphicspath{{./}{}}

\begin{document}

\title{Identifying the secondary jet in the RadioAstron image of OJ~287}

\author[0000-0001-8580-8874]{Mauri J. Valtonen}
\affiliation{FINCA, University of Turku, FI-20014 Turku, Finland}
\affiliation{Tuorla Observatory, Department of Physics and Astronomy, University of Turku, FI-20014 Turku, Finland}

\author{Lankeswar Dey}
\affiliation{Department of Physics and Astronomy, West Virginia University, PO Box 135, Willey Street, Morgantown, WV 26506
USA}



\author[0000-0003-3609-382X]{Staszek Zola}
\affiliation{Astronomical Observatory, Jagiellonian University, ul. Orla 171, 30-244 Krakow, Poland; szola@oa.uj.edu.pl}

\author[0000-0002-9331-4388]{Alok C. Gupta}
\affiliation{Aryabhatta Research Institute of Observational Sciences (ARIES), Manora Peak, Nainital -- 263001, India}

\author{Shubham Kishore}
\affiliation{Aryabhatta Research Institute of Observational Sciences (ARIES), Manora Peak, Nainital -- 263001, India}

\author{Achamveedu Gopakumar}
\affiliation{Department of Astronomy and Astrophysics, Tata Institute of Fundamental Research, Mumbai 400005, India}


\author[0000-0002-1029-3746]{Paul J. Wiita}
\affiliation{Department of Physics, The College of New Jersey, 2000 Pennington Rd., Ewing, NJ 08628-0718, USA}

\author[0000-0002-4455-6946]{Minfeng Gu}
\affiliation{Shanghai Astronomical Observatory, Chinese Academy of Sciences, 80 Nandan Road, Shanghai 200030, People's Republic of China}

\author{Kari Nilsson}
\affiliation{FINCA, University of Turku, FI-20014 Turku, Finland}


\author[0000-0002-8366-3373]{Zhongli Zhang}
\affiliation{Shanghai Astronomical Observatory, Chinese Academy of Sciences, 80 Nandan Road, Shanghai 200030, People's Republic of China}
\affiliation{Key Laboratory of Radio Astronomy and Technology, Chinese Academy of Sciences, A20 Datun Road, Chaoyang District, Beijing 100101, People's Republic of China}

\author{Rene Hudec}
\affiliation{Faculty of Electrical Engineering, Czech Technical University, 166 36 Prague, Czech Republic; rene.hudec@gmail.com}
\affiliation{Astronomical Institute (ASU CAS), 251 65 Ond\v{r}ejov, Czech Republic}

\author{Katsura Matsumoto}
\affiliation{Astronomical Institute, Osaka Kyoiku University, 4-698 Asahigaoka, Kashiwara, Osaka 582-8582, Japan; katsura@cc.osaka-kyoiku.ac.jp}

\author{Marek Drozdz}
\affiliation{Mt. Suhora Astronomical Observatory,  University of the National Education Commission,
ul.Podchorazych 2, 30-084 Krakow, Poland}

\author{Waldemar Ogloza}
\affiliation{Mt. Suhora Astronomical Observatory,  University of the National Education Commission,
ul.Podchorazych 2, 30-084 Krakow, Poland}

\author{Andrei V. Berdyugin}
\affiliation{Tuorla Observatory, Department of Physics and Astronomy, University of Turku, FI-20014 Turku, Finland}

\author{ Daniel E. Reichart}
\affiliation{Department of Physics and Astronomy, University
 of North Carolina at Chapel Hill, Chapel Hill, NC 27599, USA; reichart@unc.edu }

\author[0009-0002-2791-1341]{Markus Mugrauer}
\affiliation{Astrophysical Institute and University Observatory, Schillergässchen 2, D-07745 Jena, Germany; markus@astro.uni-jena.de}

\author{Tapio Pursimo}
\affiliation{ Nordic Optical Telescope, Apartado 474, E-38700 Santa Cruz de La Palma, Spain; tpursimo@not.iac.es}

\author[0000-0002-0712-2479]{Stefano~Ciprini}
\email{stefano.ciprini@ssdc.asi.it}
\affiliation{Istituto Nazionale di Fisica Nucleare, Sezione di Roma
``Tor Vergata", I-00133 Roma, Italy}
\affiliation{Space Science Data Center - Agenzia Spaziale Italiana, Via
del Politecnico, snc, I-00133, Roma, Italy}

\author{Tatsuya Nakaoka}
\affiliation{Hiroshima Astrophysical Science Center, Hiroshima University, 1-3-1 Kagamiyama, Higashi-Hiroshima, Hiroshima 739-8526, Japan}

\author[0000-0002-7375-7405]{Makoto Uemura}
\affiliation{Hiroshima Astrophysical Science Center, Hiroshima University, 1-3-1 Kagamiyama, Higashi-Hiroshima, Hiroshima 739-8526, Japan}

\author[0000-0002-0643-7946]{Ryo Imazawa}
\affiliation{Department of Physics, Graduate School of Advanced Science and Engineering, Hiroshima University, 1-3-1 Kagamiyama, Higashi-Hiroshima, Hiroshima 739-8526, Japan}

\author{Michal~Zejmo}
\affiliation{Janusz Gil Institute of Astronomy, University of Zielona Gora, Lubuska 2, 65-265 Zielona Gora, Poland;  mzejmo@uz.zgora.pl}

\author{Vladimir~V.~Kouprianov}
\affiliation{ Department of Physics and Astronomy, University of North Carolina at Chapel Hill, Chapel Hill, NC 27599, USA; v.kouprianov@gmail.com }

\author{James W. Davidson, Jr.}
\affiliation{Department of Astronomy, University of Virginia, 530 McCormick Rd., 
Charlottesville, VA, 22904, USA; jimmy@virginia.edu}
 
\author{Alberto Sadun}
\affiliation{ Department of Physics, University of Colorado, Denver, CO 80217, USA; alberto.sadun@ucdenver.edu}

\author{Jan \v{S}trobl }
\affiliation{Astronomical Institute (ASU CAS), 251 65 Ond\v{r}ejov, Czech Republic}


\author{Martin Jel\'{\i}nek}
\affiliation{Astronomical Institute (ASU CAS), 251 65 Ond\v{r}ejov, Czech Republic}

\author{Abhimanyu Susobhanan}
\affiliation{Department of Astronomy and Astrophysics, Tata Institute of Fundamental Research, Mumbai 400005, India}

\begin{abstract}
\noindent
The 136 year long optical light curve of OJ~287 is explained by a binary black hole model where the secondary is in a 12 year orbit around the primary. Impacts of the secondary on the accretion disk of the primary generate a series of optical flares which follow a quasi-Keplerian relativistic mathematical model. The orientation of the binary in space is determined from the behavior of the primary jet. Here we ask how the jet of the secondary black hole projects onto the sky plane. Assuming that the jet is initially perpendicular to the disk, and that it is ballistic, we follow its evolution after the Lorentz transformation to the observer's frame. Since the orbital speed of the secondary is of the order of one-tenth of the speed of light, the result is a change in the jet direction by more than a radian during an orbital cycle. We match the theoretical jet line with the recent 12 $\mu$as-resolution RadioAstron map of OJ~287, and determine the only free parameter of the problem, the apparent speed of the jet relative to speed of light. It turns out that the Doppler factor of the jet, $\delta\sim5$, is much lower than in the primary jet. Besides following a unique shape of the jet path, the secondary jet is also distinguished by a different spectral shape than in the primary jet. The present result on the spectral shape agrees with the huge optical flare of 2021 November 12, also arising from the secondary jet.     
\end{abstract}

\section{Introduction} \label{sec:introduction}
\noindent
The low redshift \citep[$z=0.306\pm0.001$,][]{1985PASP...97.1158S,2010A&Ap...516A..60N} blazar OJ~287 has attracted a lot of attention due to the availability of a 136 year-long optical light curve. Its prominent features are large flares. The periodicity of the flares was first noticed by  Aimo Sillanp\"a\"a in 1982, and since then seven more flares have been observed that fit the same pattern. All except one of these (the 1984 flare) were predicted in advance, the most prominent being the 2019 flare which came only 90 minutes later than anticipated \citep{2020ApJ...894L...1L}. One more flare happened since then, in 2022, but because of instrumental limitations it was caught only at a pre-stage \citep{2023MNRAS.521.6143V,2024RNAAS...8..276V}. At the same time, more flares were discovered in historical photographic plate studies so that only eight of the expected 26 flares remain unconfirmed \citep{2013A&Ap...559A..20H}. All unconfirmed ones are due to lack of known photographs at the expected epochs.

 The model is based on the idea that the system consists of a supermassive black hole binary of unequal masses. Every time the secondary black hole impacts the accretion disk of the primary, hot bubbles of gas are pulled out of the disk, and it is the bremsstrahlung radiation of these bubbles that reveal the orbit of the binary \citep{1998ApJ...507..131I}. The radiation peak arises when the bubble becomes suddenly transparent, and the relevant time scale is the light-travel time through the bubble. In practice we are discussing days, depending somewhat on how close to the primary the impact happens. Using the timings of ten flares \citet[][hereafter, Paper I]{2018ApJ...866...11D}  found an exact mathematical solution to the flare timing problem, with nine accurately determined parameters. Among the parameters are the primary mass $m_1 = 18.35\pm0.05 \times 10^9 M_{\odot}$, the secondary mass $m_2 = 150\pm10 \times 10^6 M_{\odot}$, primary Kerr parameter $\chi_1 = 0.38\pm0.05$, orbital eccentricity $e = 0.657\pm0.003$, and orbital period (redshifted) $P = 12.06\pm0.01$ years. One of the parameters in Paper I is the so called tail term in gravitational radiation. Since it is calculable, it is not really a free parameter, and therefore the actual number of parameters is eight, and they require nine flares for the complete solution of the problem.
 
However, the dominant optical emission of OJ~287 comes almost always from the jet of the primary black hole. The overall light curve is well explained by the variations of the accretion flow, induced by the presence of the secondary \citep{1997ApJ...484..180S}. The accretion flow variations not only explain the past light curve since year 1900, but they have also made successful predictions, most notably the 2020 April-June superflare which was intensively observed \citep{2020MNRAS.498L..35K}. The relevant time scale for the accretion-flow flares is the half-period of the innermost stable orbit in the accretion disk, about 50 days \citep{2013MNRAS.434.3122P}.

The radio light curve requires additionally modeling of the direction of the jet, leading to changes in Doppler boosting, as modulated by the binary motion. This explains the overall radio variability structure in the data that have been collected since early 1970's \citep{2006ApJ...646...36V}. There is a four year delay between the changes in the initial jet and the radio variability peaks. This is understood as a delay of transmission of information from the center to distant parts of the jet. The delay is frequency dependent: the 4-year delay applies to high frequencies (37 GHz) while at low frequencies the delay is longer \citep{2021MNRAS.503.4400D}.

Another long data set exists concerning the optical polarization of OJ~287, beginning also in the early 1970's. The optical polarization angle appears closely connected to the projected direction of the innermost jet in the sky. It is similarly modulated by the binary motion \citep{2012MNRAS.421.1861V,2023ApJ...957L..11G}. Since the optical frequencies are very much higher than radio frequencies, the delay in transmission of information to the optically radiating part of the jet is short, and the variability in optical reacts to changes at the center practically without delay \citep{2021Galax..10....1V}.

 \begin{figure*}
 \centering
\includegraphics[width=0.40 \textwidth,angle=0] {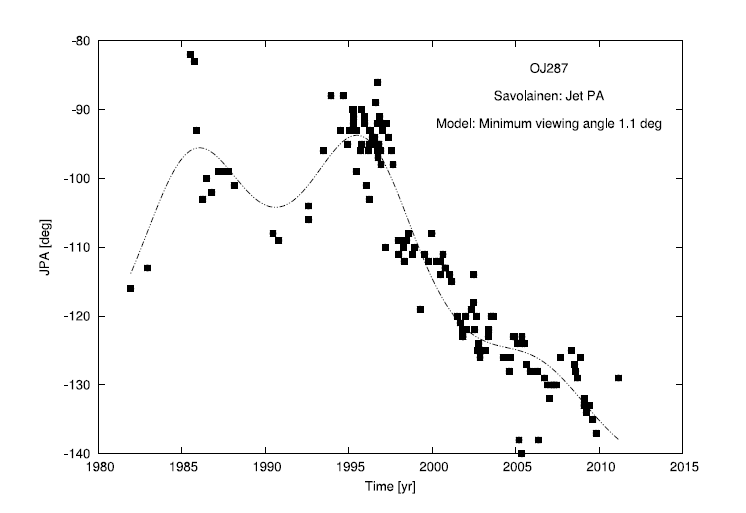}
\includegraphics[width=0.40 \textwidth, angle=0] {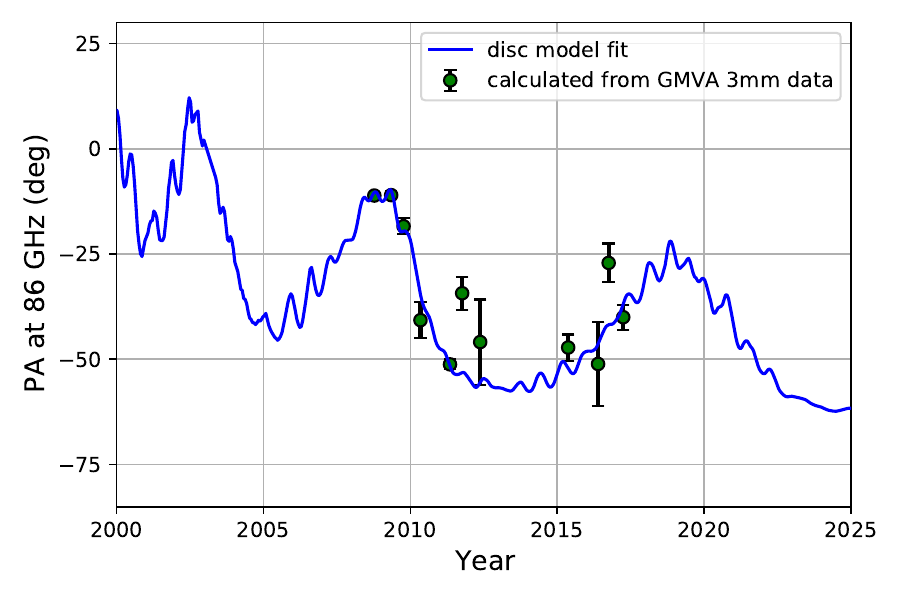}
\caption{The position angle of the radio jet in observations at 5 GHz and 8 GHz (points) from 1981 to 2010 and the theoretical line in the binary model \citep{arXiv.1208.4524} (left). Subsequent observational points at 86 GHz have continued to remain more or less on the theoretical line \citep{2021MNRAS.503.4400D} (right). Preliminary data from 2017 to 2021 indicate that the jet continues to follow the model (G.-Y. Zhao, private communication).}
\label{fig:3}
\end{figure*}

While the orbit is accurately determined, its orientation in space cannot be determined without high spatial resolution. The required resolution is obtained only at radio frequencies. Fortunately, as recent observations have shown, the innermost accretion disk in nearby AGN is often prominent at radio waves, and thus the position of black holes can be determined even if the resolution is not great enough to detect the tell-tale ring around the black holes \citep{Goddi_2021_ApJL_910_L14,2022ApJ...940...65O,2023Natur.616..686L}.

The primary jet of OJ~287 is resolved in observations. The observations of the jet position angle (PA) go as far back as the 1981 \citep{1987ApJ...323..536R}. It is possible to model the jet behavior simply by calculating the influence of the binary motion on the inner accretion disk of the primary, and to assume that the jet comes out perpendicular to this disk  \citep{2011IAUS..275..275V}, an assumption which is justified by magnetohydrodynamic simulations  \citep{2018MNRAS.474L..81L}. As Figure 1 shows, the data fitted well with the model up to year 2011 when the model was first published. Besides explaining the 12-year periodic structure, it predicted a large PA jump in the jet at about this time. The jump was seen \citep{2012ApJ...747...63A} and was referred to as "erratic jet wobbling", as the authors were not aware of the predicted behavior of the jet. It was later shown that the jump arises from the jet passing essentially through the line of sight which also causes a 180-degree flip in the position angle \citep{2012MNRAS.421.1861V,2013A&A...557A..28V}.

\cite{2021MNRAS.503.4400D} (Paper II in the following) improved the theoretical model and took account of the jet PA observations at three radio frequencies. The new model used a hydrodynamical version of the particle code to model the inner accretion disk. Its orientation varies in a cyclical manner with the 12 year period, and this is reflected in the PA observations. The best match between theory and observations arise at the highest observing frequency of 86 GHz. The observing frequency matters, as the highest frequency data refer to the innermost jet, according to the model.  

Using this model it is possible to uniquely solve for the orientation of the binary system in space. In this way, we have complete information on the OJ~287 binary orbit, both on its intrinsic parameters and on its projection onto the sky plane. Paper II also calculates the position of the secondary black hole in the sky with respect to the primary at the apocenter of the orbit, and the corresponding secondary jet direction at the same epoch (in 2013).

\begin{figure*}
\hspace{-1cm}
 \centering
 \includegraphics[scale=0.2]
{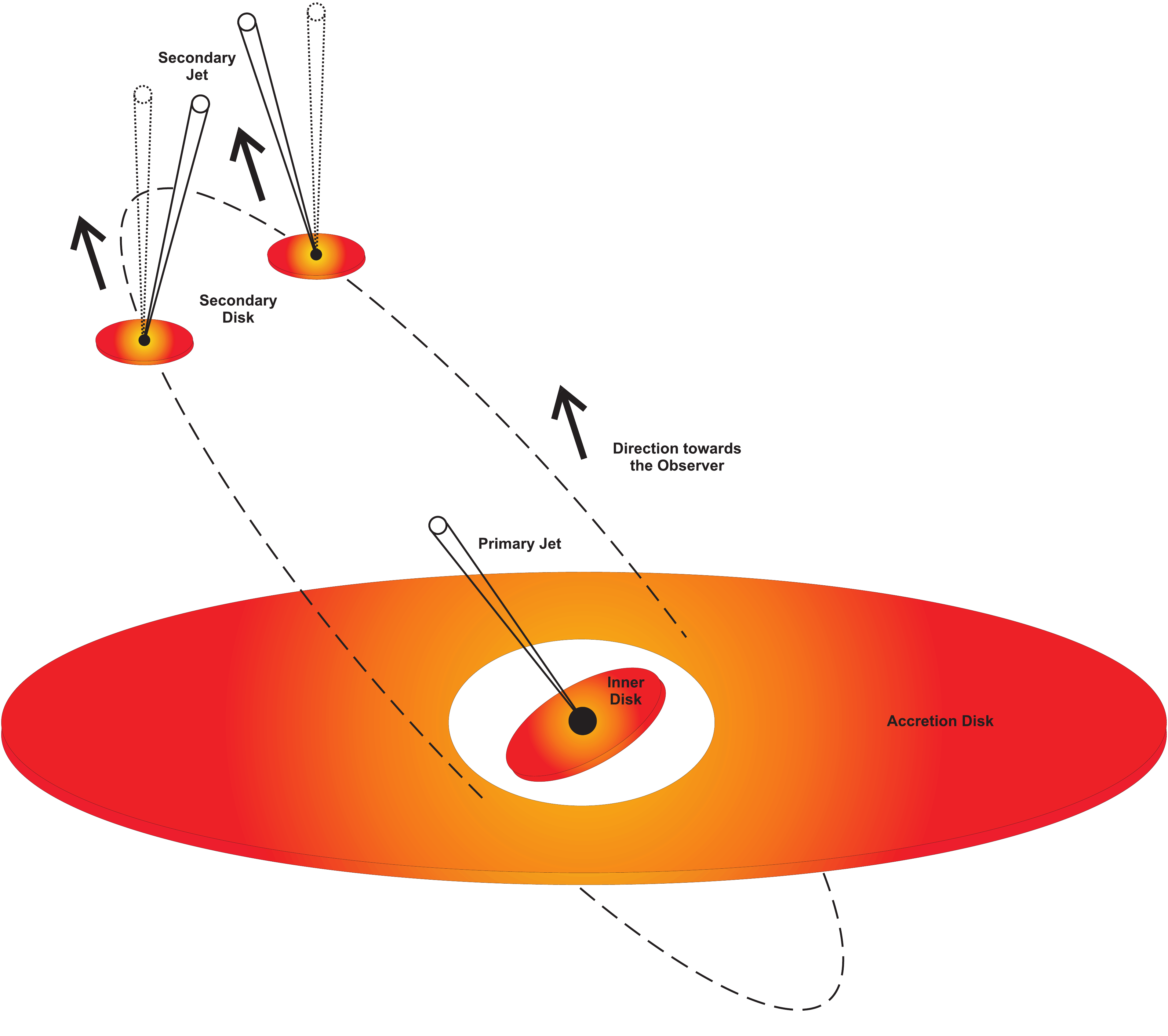} 
   \caption{An illustration of the proposed astrophysical system of OJ287. The primary black hole lies at the focal point of the eccentric orbit which has been determined in Paper I. The primary black hole is surrounded by an accretion disk. Due to spin-orbit interaction, the central part of the disk out to a few Schwarzschild radii is slanted relative to the main disk \citep{1975ApJ...195L..65B}. There is a smooth transition from the inner disk to the outer disk (unlike in the figure). The primary jet starts out along the axis of the disk. The time evolution of the axis of the inner disk is calculated in Paper II. The secondary black hole is shown at two different orbital phases. Its accretion disk is parallel to the main accretion disk, and its jet therefore starts out perpendicular to the main disk. However, after Lorentz transformation to the observer's frame, the secondary jet is seen tilted in the direction of the orbital motion, and thus its direction changes constantly. Here the orbital motion is counter-clockwise.} 
    \label{fig 2}
\end{figure*}

As said above, Paper II models the behavior of the inner accretion disk of the OJ 287 binary using the disk simulation code of \cite{https://doi.org/10.1143/PTPS.195.48}, in a manner explained in \cite{2013ApJ...764....5P}. The inner disk is found to closely follow the spin direction of the primary black hole except that its axial direction varies somewhat faster than the spin direction. Assuming that the direction of the primary jet is perpendicular to the inner disk, the jet direction is determined at every moment of time. See Figure 2. This direction shows up in VLBI observations as the temporal evolution of the projection of the radio jet in the sky. Data from three radio frequencies were used, with the resolution of the jet increasing with frequency. In this way a snapshot of the jet at three different distances from the center were taken. This information was enough to solve the main parameters of the problem uniquely: the jet viewing angle at some initial time, and at all subsequent times, and the direction of the major axis of the binary orbit in sky projection. The parameters at the three frequencies vary slightly, indicating that the jet has a helical structure.

In Figure 3 (left panel) we show a detail of Paper II: the primary black hole is at the center, and the secondary black hole is at the position angle of about 35 degrees, at the distance of 20 $\mu$as from the primary black hole. This is when the secondary was at its recent apocenter point in 2013. Since the binary orbit is expected to be seen more or less edge-on, the secondary will move back and forth in a nearly straight line connecting the two black holes, going from side to side over many orbits.

The projected jets coming from the black holes are given by lines: the primary jet direction varies as a function of time, with the color code referring to different moments of time. The primary jet position angle is around 315 degrees, or $-45$ degrees as we label it in this paper. The direction of the secondary jet at 2013, when the secondary has no transverse motion relative to us, is shown as a dashed line. Since the secondary is in constant motion, the secondary jet direction is a snapshot and not generally applicable.

Since the calculation of Paper II, a sub-parsec radio map was made available by \citet[][Paper III in the following]{2022ApJ...924..122G}, capable of detecting both black holes as well as the secondary jet, if they are bright enough. On the right hand panel of Figure 3 we show a detail of this map (provided kindly by J.L. Gomez) which has the unprecedented resolution of 12 $\mu$as in the direction of the line connecting the two black holes of the left hand panel. They are the two black dots, at the position angle of about 35 degrees with respect to each other, on the left hand panel. The region shown here is thought to include the primary black hole, at the center, represented by a point source. Point sources appear elliptical rather than circular in shape because of technical limitations of the space VLBI interferometry. Two other point sources are also seen. The comparison with the theoretical map of Paper II suggests a possible identification of the two other point sources with the secondary black hole and its innermost jet, respectively. This high resolution was achieved by using a single space antenna (RadioAstron) together with the global VLBI network. The RadioAstron antenna was 15 Earth diameters away from us, giving the maximum baseline of 190,000 km. The RadioAstron orbit has an apogee that is even longer, but the detections were registered at these baselines. Since we had only a single space antenna, the high resolution was achieved only in one direction, but fortunately, this is the direction most interesting to us. The higher resolution in one direction is not only caused by the limitation to one antenna, but also by the projection of the orbit onto the plane of the sky at this position angle.

Paper III discussed the map from the point of view of a single jet interpretation which is certainly valid, and at the time when no evidence for the visibility of the secondary jet had yet shown up, was the only reasonable thing to do.

 However, \citet[][Paper IV in the following]{2024ApJ...968L..17V} point out that the rapid flare in OJ~287 on 2021 November 12 originated most likely from the secondary jet. Here the time scale is about 4 hours \citep{2013ApJ...764....5P}, the time scale expected from the period of the innermost stable orbit in the accretion disk of the secondary black hole. The main argument is the size of the radiating region, determined from the variability time scale measured by \citet{2024ApJ...960...11K} which is orders of magnitude too small for the primary jet, even below the Schwarzschild radius of the primary black hole, but in agreement with the expected cross-section of the secondary jet. Another argument comes from the spectral energy distribution, which peaks at a much higher frequency during the flare than outside it. Therefore we may ask if the secondary jet might also show up in the high-resolution radio map, even though the RadioAstron radio map was taken about a year after the expected bright optical flare in 2013 (which was not observed due to insufficient monitoring coverage, see Paper IV). Paper IV also discusses the polarization properties of the flare and concludes that the rapid flare is highly polarized but along a different axis than the normal background emission, leading to a reduction in the overall degree of polarization. This behavior is expected if we are dealing with optically thin synchrotron radiation from two independent sources. In this paper, we assume this to be the case.
 
The cosmological model adopted in Paper III has $\Omega_M=0.27, \Omega_{\Lambda}=0.73$ and $H_0=71~ {\rm km s}^{-1} {\rm Mpc}^{-1}$, and we use it also in this paper. The angular scale of 20 $\mu$as then corresponds to 18,481 au, which is practically the same as the apocenter distance of the binary orbit. In calculation of speeds, we note that the intrinsic time intervals at OJ~287 are shorter than the corresponding interval observed by us by a factor of 1.306. 

\section{Secondary core}
\noindent  
Supposing that the core of the primary jet is the component C1a of Paper III, then Paper II places the secondary black hole exactly on the first knot in the jet  (see Paper III). This knot would be about one contour level, or factor of three, fainter than component C1a. See  Figure 3. Nothing else is known about this knot, so we may ask if it is reasonable to expect the secondary core to have this radio brightness.

\begin{figure*}
\centering
\includegraphics[width=0.36 \textwidth, angle=0] {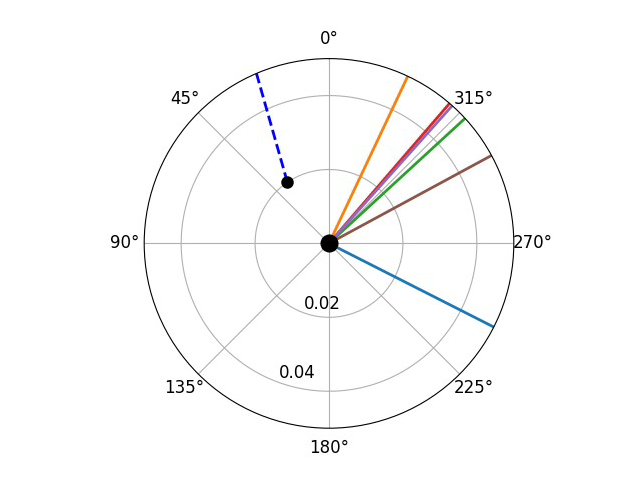}
\includegraphics[width=0.32 \textwidth, angle=0] {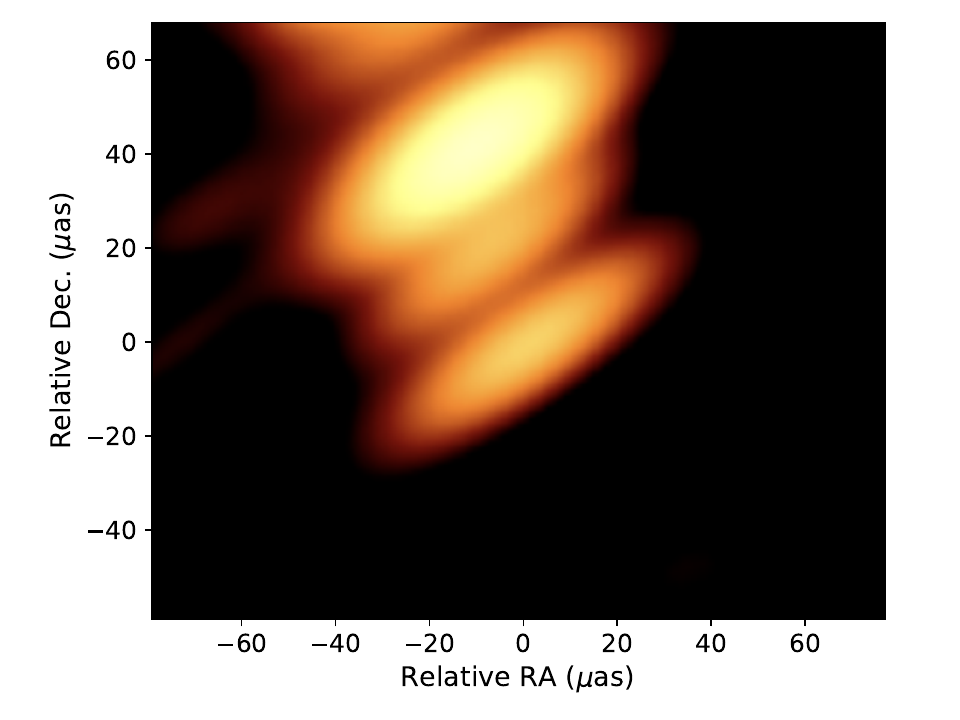}
\caption{Left panel: The theoretical model of Paper II, after fitting to the radio jet observations. The primary black hole is at the center and the secondary black hole is at the position angle of about 35 degrees. The blue dashed line is the initial direction of the secondary jet, while solid lines represent the primary jet at different times. The 2014 jet direction should be close to the green line. Right panel: A detail of the RadioAstron map of Paper III in the same scale and orientation as the theoretical map. If the central component corresponds to the primary black hole, then the next one upwards marks the secondary black hole, and the highest component represents a knot in its jet. The elongation of the individual components is not real, but is a reflection of the beam shape.}
\label{fig:2017_and_historical_fades}
\end{figure*}

In Paper I the spin of the primary has been determined to have a relatively low value \citep{2016ApJ...819L..37V}, while we have good reasons to believe that the secondary black hole spins near its maximum speed (see Paper IV). This high spin would be expected because the secondary receives and accretes a new dose of gas at every disk crossing, always with the same direction of the specific angular momentum.

In a jet whose power is extracted from the black hole's spin, via the \textit{Blandford-Znajek} mechanism \citep{1977MNRAS.179..433B}, the jet luminosity $L_{j}$ is estimated as a function of the normalized spin $J$, the black hole mass $m$ and the accretion rate $\dot{m}$ (in the Eddington units) as described by  \citet{1997MNRAS.292..887G}:

\begin{equation}
L_{j} \sim m^{1.1} \dot{m}^{0.8} J^2.                 
\end{equation}

Using the parameters from Paper I and from \citet{2019ApJ...882...88V}, the primary spin is $J\sim 0.38$ and the primary mass is $m\sim 122$ times greater than the secondary mass, while for the secondary, $J \sim 1$. Thus putting $\dot{m} = 1$ for the secondary, a likely value for temporary jets which are created by a sudden burst of accreting matter,  and $\dot{m}\sim 0.08$ for the primary makes the secondary jet $\sim 1/4$ of the luminosity of the primary. An Eddington accretion rate onto the secondary is reasonable at the disk crossing time when the Roche lobe of the secondary has been recently supplied by gas of the primary disk. The calculated luminosity ratio is not far from the luminosity ratio $\sim 1/3$ that is obtained by comparing the two putative cores in the RadioAstron map. Note that the resolution of the RadioAstron map is such that we are not anywhere close to seeing the inner accretion disk rings around the two black holes, but will be looking at the inner parts of the jets, and thus the jet luminosity ratio is relevant.

There is no other proof that the second knot in the jet is actually the radio core of the secondary. We have to wait for new maps at the 10 $\mu$as resolution or better at different frequencies in order to make further deductions about its nature. The Radioastron system has not been available since 2019, and no comparable missions are currently planned. Even though the core will not be seen in future VLBI experiments for some time to come, the secondary jet should be easily resolved, as we will discuss below.

\section{Aberration}
\noindent

We calculate the track of the secondary jet on two assumptions: (1) the jet starts out perpendicular to the secondary accretion disk in the frame of the secondary black hole; and (2) that the jet is ballistic. The secondary accretion disk is taken to be parallel to the primary accretion disk, since both have the same origin of angular momentum. In the observer's frame, the jet is bent in forward direction with respect to the motion of the secondary due to aberration. Suppose that in the reference frame of the observer the source is moving with the bulk speed $v$ at an angle $\theta$ relative the line from the observer to the light source at the time when the light is emitted. Then using the Lorentz transformation, the observer sees the aberration of light $\theta_a$ given by 
\begin{equation}
\tan(\theta_a/2) = [(1+v/c)/(1-v/c)]^{1/2} \tan(\theta/2).                 
\end{equation}
To estimate the magnitude of aberration, we note that the typical orbital motion calculated in Paper I is $\sim 0.1c$, where $c$ is the speed of light. If the jet moves close to $c$, then the magnitude of aberration is $\sim 0.1$ radians. With the viewing angles found in Paper II,  $\Theta_{\rm obs} = 4.70\pm0.25^{\circ}$ and $\Phi_{\rm obs}=72.0\pm0.5^{\circ}$, this aberration is projected in the sky plane as an angle of $\sim 1$ radian. Thus, aberration cannot be ignored in this problem.

The polar angle $\Phi$ is measured in the plane of the disk and the angle $\Theta$ is the distance from the disk axis. The orbital motion takes place in the plane defined by $\Phi$ = $\Theta$ = $0^{\circ}$. See Paper II for the definition of the coordinate system.

The speed of the secondary perpendicular to the line of sight was calculated in Paper I, and its approximate values may be calculated from the coordinates given in Table 2 of Paper IV. Note that the values in the table are in astronomical units and in Earth years. The conversion factor from au/yr at Earth to the speed of light units at OJ~287, including the redshift factor 1.306, is $2.069 \times 10^{-5}$. Here we use exact coordinate values for higher accuracy. As aberration is a function of time, we give its values from 2014.32, the approximate time of the RadioAstron map, backward in steps of 1/3 of a year. The last few backward steps in 2007 are at shorter time intervals because the speed changes rapidly. The results are shown in columns 1 and 2 of Table 1.

\begin{table}
\caption{The effect of aberration on the secondary jet.}   
\begin{tabular}{lcccccr}
\hline
         $Epoch$ & $Aber$ & $\Delta R$ & $PA$ & $PA_c$ & $r$ \\
         year & deg & $\mu$as  & deg & deg & $\mu$as\\
\hline         
     2007.80  &     2.3    &     23.6  &  $-14$ & $-19$ & 365\\
     2007.87  &     4.1    &     23.4  &  $-25$ & $-30$ & 361\\
     2007.90  &     4.7    &     23.3  &  $-28$ & $-33$ & 359\\
     2007.94  &     5.4    &     23.1  &  $-31$ & $-35$ & 357\\
     2008.00   &    6.2    &     22.7  &  $-34$ & $-38$ & 354\\ 
     2008.30   &    8.0    &     20.5  &   $-39$ & $-43$ & 337\\
     2008.67   &    7.9    &     17.6  &  $-39$ & $-42$ & 316\\
     2009.01   &    7.3    &     15.1  &  $-37$ & $-40$ & 297\\
     2009.30   &    6.7    &     13.0  &  $-36$ & $-37$ & 281\\
     2009.67   &    5.9    &     10.7  &  $-33$ & $-35$ & 260\\
     2010.01   &    5.3    &     8.8  &  $-31$ & $-32$ & 241\\ 
     2010.31   &    4.7    &     7.3  &   $-28$ & $-28$ & 225\\
     2010.67  &     4.1    &     5.8  &  $-25$ & $-26$ & 204\\
     2011.01   &    3.6    &     4.5  &  $-23$ & $-24$ & 185\\ 
     2011.31   &    3.1    &     3.5  &  $-20$ & $-20$ & 169\\
     2011.68   &    2.6    &     2.4  &   $-16$ & $-16$ & 148\\
     2012.00   &    2.1    &     1.7  &   $-13$ & $-13$ & 130\\
     2012.32   &    1.7    &     1.1  &  $-9$ & $-9$ & 112\\
     2012.68   &    1.2    &     0.6  &  $-5$ & $-5$ & 92\\
     2013.00   &    0.8    &     0.2  &  $-1$ & $-1$ & 74\\
     2013.31   &    0.4    &     0.1  &  3 & 3 & 57\\ 
     2013.68   &    0.0    &     0.0  &  8 & 8 & 36\\
     2014.01   &   $-0.5$    &     0.1  &  14 & 14 & 17\\ 
     2014.32   &   $-0.9$    &     0.3  &   19 & 19 & 0\\
\hline
\end{tabular}
\label{TabAllRData}
\end{table}

In this model the jet starts from the secondary black hole and consequently its position in the sky changes with respect to the primary. Column 3 of Table 1 provides the amount of the change with respect to the 2014.32 position in $\mu$as units. The direction of the change is toward the primary which can be considered stationary for all practical purposes.

In column 4 of Table 1 we show the calculated position angle (PA) of the jet in the sky at different epochs. The jet PA in the sky that corresponds to each value of aberration is calculated from Eq. 4 of Paper II:
\begin{widetext}
\begin{equation}
    PA = \arctan \left( \frac{\cos{\Theta_{\rm obs}} \sin{\Theta_{\rm jet}} \cos{(\Phi_{\rm jet} - \Phi_{\rm obs})} - \sin{\Theta_{\rm obs}} \cos{\Theta_{\rm jet}}}{\sin{\Theta_{\rm jet}} \sin{(\Phi_{\rm jet} - \Phi_{\rm obs})}} \right) + \Delta PA .
\end{equation}
\end{widetext}

As mentioned above, the orbital plane is defined by $\Phi$ = $\Theta$ = $0^{\circ}$ (Paper II). Thus for the jet $\Phi_{\rm jet}$ = $0^{\circ}$ and  $\Theta_{\rm jet}$ is given by the aberration angle ({Column 2 in Table 1). The direction of the line of sight is specified by $\Theta_{\rm obs} = 4.7^{\circ}$ and $\Phi_{\rm obs}=72^{\circ}$, according to the solution of Paper II. $\Delta PA$ defines the orientation of the orbital plane in the sky. In Paper II it varies depending on what frequency is used to determine it because of the twisting of the jet; we use a typical value $\Delta PA = - 82^{\circ}$ which is the median value at the three frequencies in the spin model.

Since the position of the secondary changes, the PA's have to be corrected to the coordinate origin at the time when the ballistic jet leaves the secondary. Using the values in Columns 3 and 4 we calculate the corrected PA$_c$s and give them in Column 5. These are the values to be compared with radio maps.

The last column in Table 1 gives the radial coordinates of the ballistic jet at different times. Here we encounter the only free parameter of the problem: the apparent speed of the ballistic jet in the sky $\beta_T$, normalized to $c$. The numbers in Column 6 have been calculated assuming that $\beta_T=1$. If it is different, the scales will be stretched or compressed accordingly. For the typical viewing angle $\Theta_{\rm obs} = 4.7^{\circ}$ , this corresponds to the actual speed $\beta=0.927$ with respect to $c$, the relativistic Lorentz factor $\gamma \sim 2.7$ and the Doppler factor $\delta\sim5$. The transverse speed is rather low in comparison with other sources of well developed jets \citep{2020AA...641A..40V}, more in line with relativistic jets observed in tidal disruption events \citep{2020NewAR..8901538D}. This may not be surprising considering the intermittent nature of the secondary jet.

When we overlay the jet-line with the RadioAstron image, we see that the secondary jet follows components C1b to C2 and J6. Component J5 belongs to the primary jet in this description, and it goes further to components J4 - J1 (see Paper III). Both jets turn about 90 degrees in all, due to small angle variations and projection effects (Paper II). 

\begin{figure*}
\centering
\includegraphics[width=1.0 \textwidth, angle=0] {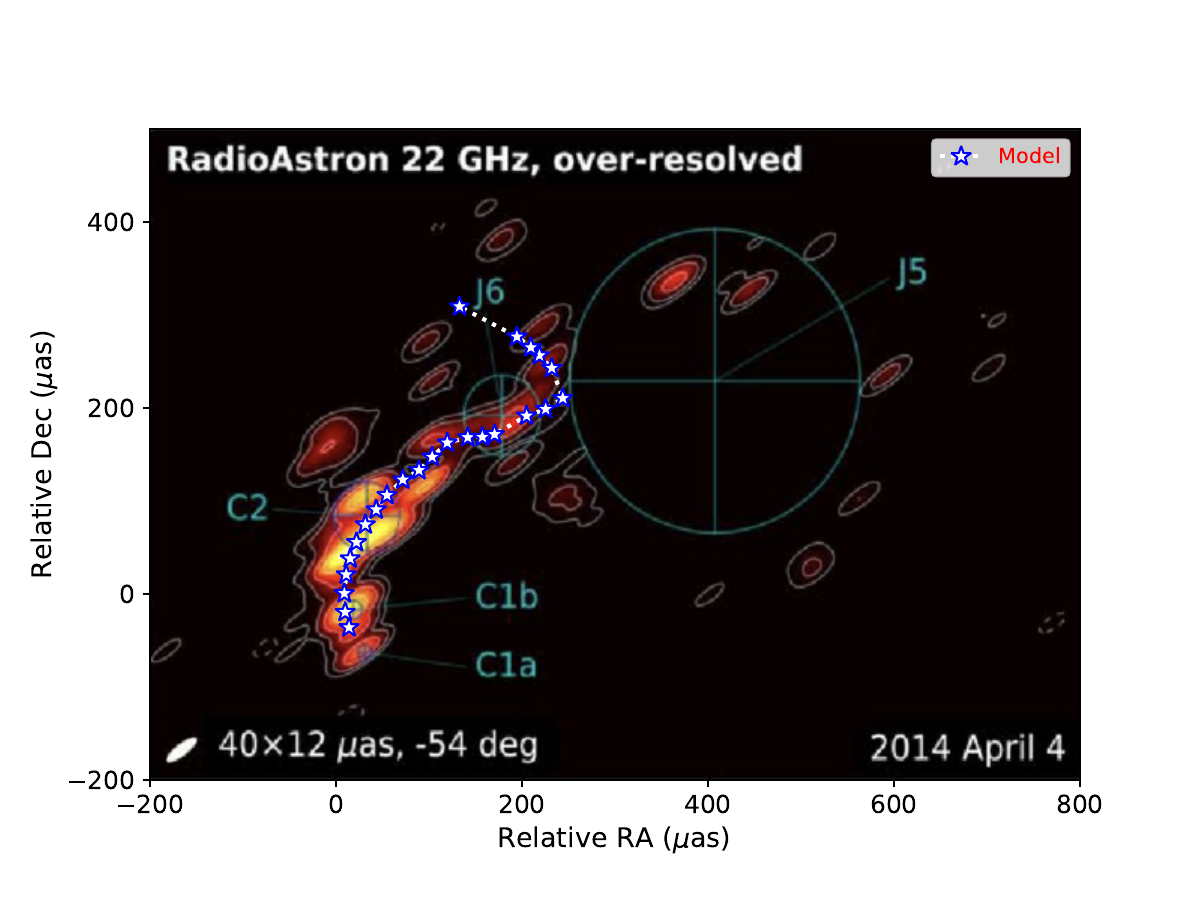}
\caption{The image of OJ~287 at the record-breaking resolution of 12 $\mu$as, achieved in space VLBI when the RadioAstron telescope was 15 Earth diameters away from the ground based telescopes (a distance of about 190,000 km, comparable to about half of the semi-major axis of the Moon orbit). The present primary jet direction is at about $-45^{\circ}$ degrees, i.e. it goes through the component J5. The path of the primary jet between the primary core C1a and J5 is not known, and has to be determined by future high resolution observations. The highest resolution image agrees with the secondary jet, starting from the knot between components C1a and C1b toward the position angle of about $+15 ^{\circ}$ degrees. The two innermost components at the beginning of the jet lie at positions expected for the two black holes at the time of the observations in 2014 (see Fig. 1). Superposed on the map is the jet line from Table 1 (star symbols). For the original RadioAstron map, see Paper III.}
\label{fig 4}
\end{figure*}
\section{Discussion}
\noindent
Besides the steady jets, it has been realized in recent years that new jets can suddenly appear, for example, in X-ray binaries and during tidal disruption events \citep{2012A&A...538A...5K,2020NewAR..8901538D}. Therefore, it is not excluded that new jets may appear also in the OJ~287 system. According to numerical simulations, the trigger for the new jet  coming from the secondary black hole would be its crossing through the accretion disk of the primary \citep{2013ApJ...764....5P}. It should be short-lived and appear about twice per orbital period. Figure 2 illustrates the binary black hole model.

What happens to the jet of the secondary black hole when it goes through the accretion disk of the primary? We may assume that the jet is anchored to the innermost part of the accretion disk, say, within 3 Schwarzschild radii of the black hole \citep{1977MNRAS.179..433B} (BZ model), \citep{1982MNRAS.199..883B} (PB model). Then the question is what happens to this part of the disk during the transit. In order to compare the physical quantities in the primary and secondary disks, we may use the standard scaling laws for magnetic accretion disks \citep{1984ApJ...277..312S}. 

The density $n$ scales as

\begin{equation}
n \sim m^{-0.8} \dot{m}^{-0.4} r^{-0.6} (1-r^{-0.5})^{-0.4},             \end{equation}

while the magnetic flux density $B$ scales

\begin{equation}
B \sim m^{-0.4} \dot{m}^{-0.8} r^{-1.8} (1-r^{-0.5})^{0.8},                 \end{equation}

where $m$ is the mass of the black hole, $\dot{m}$ is accretion rate in units of the critical rate , $r$ is the distance in units of the Schwarzschild radius of the black hole \citep{2019ApJ...882...88V}. In order to calculate the ratios of $n$ and $B$ in the two disks at the relevant distances from their central black holes, we put $m=122$ (the mass ratio of the two black holes), $\dot{m}=0.08$ (the accretion rate ratio), $r=10$ (the distance ratio from each black hole, when we look at the primary disk about ten times further from the center than the secondary disk, in Schwarzschild units), we get $n\sim0.010$ and $B\sim0.012$. It means that the secondary disk has about 100 times the density of the matter that it is passing through, and its magnetic field is also about 100 times stronger than the field in the surrounding medium. Furthermore, the speed of the wind that the accretion disk reaches is only about 1/3 of the orbital speed in the disk.

It is obvious that the central part of the smaller black hole disk is hardly aware that it is traversing through the bigger disk; the latter is so tenuous and weakly magnetized that the influence must be minimal. Primarily, we expect that the disk would be somewhat puffed up with respect to the standard disk, since the total surface density  (surface density is the disk thickness times the density where the thickness is proportional to $m\times\dot{m}$ \citep{1984ApJ...277..312S}) of the big disk is about 10 percent of the surface density of the secondary disk in the standard model, enough to affect the vertical velocity dispersion in the disk. The disk crossing becomes apparent to the small disk only after the gas from the primary disk starts to flow on top of the small disk, in the manner calculated in detail by \cite{2013ApJ...764....5P}. In contrast, the tenuous bigger disk is strongly shocked and displays phenomena that we described above. The amount of gas pulled out of the disk, about $16 M_{\odot}$ per impact, is so small, that it plays no detectable role in the overall dynamics of the binary system \citep{2013ApJ...764....5P}.

In a seminal study on 3D GRMHD model of magnetized accretion onto a spinning black hole by \cite{2011MNRAS.418L..79T}, the initial condition is a torus of inner radius 15 Schwarzschild radii with a density maximum at 34 Schwarzschild radii, which are about the same scales that we are studying in the smaller tidally truncated disk. 

One can see from their Figure 1 that accretion and a strong jet power are sustained well above $10^4$ dynamical times, up to at least $3\times10^4$ dynamical times, which would correspond to about 14 years for a dynamical time of 4 hours in case of the secondary black hole. In other words, according to this study it looks like that even if the secondary disk is truncated to tens of Schwarzschild radii, a strong jet can still be maintained over the duration of the binary orbit of the secondary black hole.

Therefore, it appears that the jet of the smaller black hole is always present, in spite of the frequent collisions with the big disk. This should be confirmed by hydromagnetic simulations, but has not been done so far. It may soon become within capabilities of the current numerical codes.

There are patterns in radio source maps that are well explained by turning jets, both in extragalactic double radio sources and in Galactic sources such as SS433 \citep{1989AJ.....97..674B,2017MNRAS.471..617B}. Magnetohydrodynamic simulations produce structures which can be matched with observations when viewed from suitable angles \citep{2023ApJ...948...25N}. As we have explained above, the temporary jet in OJ~287 is likely to turn over large angles, and as far as we know from theory and observations, such turning jets are possible and even relatively common \citep{2003ApJ...594L.103G}. It is also possible that some of the X-shaped radio sources have intrinsically two jets, creating two misaligned pairs of radio lobes, rather than possessing a single turning jet \citep{1992ersf.meet..307L}.

We would like to emphasize that the model discussed here is unique in that it explains at the same time the 136 yr optical light curve and the 40 yr set of radio jet observations in OJ~287. Even though one could imagine many different scenarios in the framework of Figure 2, only a very well defined scenario satisfies the observational requirements. Other wavelength regions do not have anything comparable to these data sets in length and number of observations. In a few decades we will be also able to add e.g. X-ray or Gamma-ray light curves to the discussion with a similar weight.

With this radio emission taken to be split between the two jets, we find that there is a major spectral difference between them. The primary jet peaks at low frequencies while the secondary jet has its maximum power at high radio frequencies. The peak contours of component C2 mark the position where we expect the jet to have propagated after receiving a strong boost from the disk crossing of 2013. It is interesting that the inner jet is brightest just where it is expected to be, at the location marking this boost. In the optical region, the secondary jet flare is differentiated from normal flares of the same size such that the peak of the energy distribution is at a much higher frequency than in the normal emission (Paper IV).

Another prominent difference between the jets is in the transverse motion $\beta_T$. From Paper II we find that the motion from the core to the component J5 takes place with $\beta_T\sim4$, while the present overlay confirms that in the secondary jet $\beta_T\sim1$. Since the projection factors are rather similar for both jets, we conclude that the secondary jet is also intrinsically slower. The values of $\delta$ are $\sim 5$ and $\sim 10$, (corresponding to $\gamma \sim 2.7$ and $\gamma \sim 5.5$) in the secondary and primary jets, respectively. Applying Table 6 of Paper III to components J5 and J6 we find that the equipartition magnetic field is about 20 times higher in the secondary jet than in the primary jet; this, as well as the lower $\delta$ in the secondary contribute to the higher turnover frequency in the secondary jet than in the primary jet \citep{1981ARA&Ap..19..373K}. 

 The secondary jet line will swing back and forth around the dotted line of Figure 1 in phase with the 12 year orbital period. It should be easy to verify this ``wagging of the tail" of the secondary black hole, when we have new epochs of observations with 40 $\mu$as resolution or better. We should find the aberration tail in quite a different place and of a different shape every time we observe the secondary jet. Additionally, during the swing of the tail to the left (to the East in the sky) the viewing angle increases, and that should make the jet appear weaker. It may also be important that half of the time the secondary is behind the disk, and that could have an effect on the jet. The 2014 RadioAstron epoch was in many ways fortuitously optimal for the detection of the secondary jet.

 The column 6 of Table 1 (the distance of the "stars" from the primary in Figure 4) was calculated assuming uniform motion in the sky of the plasmons in the ballistic approximation. We have also looked in the possibility that the motion becomes non-uniform in the sky-plane and even reverses back due to changes in the viewing angle of the secondary jet. It turns out that this makes little difference to the main shape of the secondary jet, except at the very end (the last four "stars" in Figure 4), where the jet appears to turn back, instead of going North-East (as in Figure 4). A loop appears to form at the end of the jet according to this calculation. This is because of the rapid change of aberration, resulting also in the change of the viewing angle of the secondary, which again makes the transverse motion appear to slow down in the sky. Thus the distances of the last four points should really be closer to the center than in Figure 4. There are two additional parameters in this calculation, the minimum angle between the two jets and the opening angle of the primary jet. These parameters could be determined when we have at least two maps of the secondary jet, at two epochs. We have to wait until the 2022 VLBI map is released to do this.

 When the secondary is behind the primary from our viewing direction, it is possible that the two jets intermingle and complicated flow patterns arise. The last time this happened in OJ287 was around 2018. See Paper IV for the details of the orbit. The ballistic method used in this paper is obviously too simple to try to model this situation, and in fact, we are not aware that anyone has ever applied magnetohydrodynamic methods to two jets crossing each other.

The RadioAstron map produced in 2016 did not have sufficient resolution (530 $\mu$as) for this study \citep{2024A&A...683A.248C}. The 86 GHz observations at the Global VLBI were better: they produced a map of OJ~287 at 40 $\mu$as resolution on 2017.3 \citep{2022ApJ...932...72Z}. At this time, the original jet PA should have been about +68 degrees, the transverse motion $\beta_T \sim 0.5$, the Doppler factor $\delta \sim 4$ due to relatively large viewing angle of the jet of $\sim 10^{\circ}$; therefore the initial transgression of the jet to the left (East) fits inside the lowest contour levels of the component C0 in the 86 GHz map. As the jet luminosity goes roughly like $\delta^4$, the expected jet brightness is reduced at least by a factor of two as compared with the 2014 RadioAstron map. It is also possible that the secondary jet brightness fades with time since the last disk crossing: the 2014 map was produced less than a year after the crossing,  while the 2017 map was four years after, as was pointed out by \citet{2022ApJ...932...72Z}.

Quantitatively, an accretion disk cut off from mass supply should decrease in brightness exponentially. Comparing the fluxes of components C2 and C1a which are generated at one-year intervals in the model, it appears that the e-folding time, i.e. the decrease of the flux by the exponential factor $e$, is about one year. Extrapolating from there to the jet components generated between 2014 and 2017, up to three years from the observing epoch of 2017.3, we find that the flux of this part of the jet should be about 100 mJy at 22 GHz. It is within the error limits of the central component C1 with the flux $520 \pm 110$ mJy at the same frequency (Paper III). Therefore this part of the jet will blend inside the central component and is not detectable separately. However, in the larger scale, the ``eht-imaging" and ``SMILI" versions of the map found a similar inner jet shape as was seen in the 2014 RadioAstron map, as is expected if it arises from the secondary black hole.

The subsequent occasion when the ballistic jet method could be applied again was in 2022, the next disk impact from the same side as the 2013 impact. A VLBI radio map from this period of time is under preparation (G.-Y. Zhao, private communication). Note that in this case the impact takes place before the apocenter of the orbit when the secondary is still moving away from the primary, corresponding to the part of the jet in Figure 4 roughly from the sixth "star" (representing 2012.68 in Table 1) downstream. When the secondary is placed at this position, the jet starts out at the PA $\sim -20^{\circ}$, and is a little shorter than in the 2014 map. More exact comparison has to wait for the finishing of the radio map. After this the visibility of the secondary jet should disappear until 2030's when we come again to the corresponding orbital phase in OJ~287.

We emphasize that the ballistic jet model is the simpliest possible. In more advanced work, one should take into account phenomena like plasma instabilities leading to energized regions of jets. These will definitely change the detailed flux density distribution both in the primary and secondary jets of OJ~287 \citep{2020AA...641A..40V}.

\section{CONCLUSIONS}\label{sec:discuss}
\noindent
The behavior of the primary jet in OJ~287 is well understood in the same binary model that explains 18 timings of big optical flares. Until recently, the role of the secondary black hole has been distinctly "secondary", i.e. no major flares coming from it had been verified. The situation changed with the huge one-day flare, seen 2021 November 12, at 2 am UT \citep{2024ApJ...960...11K}. It is difficult to understand it as anything else than radiation from the secondary jet.

This has prompted us to take a new look at the 2014 RadioAstron radio map at 22 GHz which has a 12 $\mu$as resolution. The entire secondary jet, not just its core, should be seen. We have calculated the effect of aberration on the jet, and find that it causes a very prominent and easily recognizable pattern in the sky. This pattern is seen in the inner jet of OJ~287, and for this reason we suggest that it may indeed be the secondary jet. The shape and orientation of the secondary jet is fully determined by the dynamical model, as well as the historical record of the primary jet over the period of more than 40 yr. The only unknown quantity is the scale of the jet which depends on the average projected jet speed in the sky. By fitting the theoretical and observed radio maps, we determine also this final parameter.

This is an important parameter, as it offers a view into the internal dynamics of the secondary jet. We find that the secondary jet is relatively slow in comparison with the primary jet. Previously, in Paper III, the physical properties of the primary jet component J5 and the secondary jet component J6 (in our view) were determined. Armed with the present determination of the jet bulk flow speeds, we combine the information from paper III with ours, and find that the Doppler factor $\delta$ is about twice as high in the primary jet as in the secondary jet, while the equipartition magnetic fields $B$ are in the ratio 20:1, in favor of the secondary jet. 

One cannot exclude the possibility that the RadioAstron map has only one jet, the primary jet, as was discussed in Paper III. The path of the primary jet, as calculated in the model of Paper II, is undetermined between the central component and component J5. We see from Table 1 that the aberration angle for the secondary jet and the viewing angle of the primary jet more or less coincide between components C2 and J6. Thus the two jets could be overlapping. In the over-resolved RadioAstron map the higher surface brightness secondary jet would dominate while in lower resolution maps it would be the primary jet. The paths of the two jets would be distinctly different from the central component C1a or C1b to the knot C2.

When the resolution close to that provided by RadioAstron is achieved again, in future perhaps at 0.85 mm in the Global VLBI, in the mapping of the jet near to the disk crossings at 2022 and 2032, it would be possible to verify the ``wagging of the tail" of the secondary black hole.

\section*{ACKNOWLEDGMENTS}
\noindent
This work was begun as one of the authors (MJV) was a visitor at the Institute of Astronomy at University of Cambridge. He would like to thank Cathie Clarke and Martin Rees for making the visit possible, and for discussions on the double jet model. He also acknowledges an enlightening discussion with Benoit Cerutti on the formation of jets near spinning black holes. We also appreciate the permission of Jos{\'e} L. G{\'o}mez to use the RadioAstron maps of Paper III in our illustrations and the advance information by G.-Y. Zhao on the latest VLBI maps of OJ~287. This work was partly funded by NCN grant No. 2018/29/B/ST9/01793 (SZ) and JSPS KAKENHI grant No. 19K03930 (KM). This work was partially supported by a program of the Polish Ministry of Science under the title ‘Regional Excellence Initiative’, project no. RID/SP/0050/2024/1. SC acknowledges support by ASI through contract ASI-INFN 2021-43-HH.0 for SSDC, and Instituto Nazionale di Fisica Nucleare (INFN). RH acknowledges the EU project H2020 AHEAD2020, grant agreement 871158, and internal CTU grant SGS21/120/OHK3/2T/13. MFG is supported by the National Science Foundation of China (grant 12473019), the National SKA Program of China (grant No. 2022SKA0120102), the Shanghai Pilot Program for Basic Research-Chinese Academy of Science, Shanghai Branch (JCYJ-SHFY-2021-013), and the China Manned Space Project with No. CMS-CSST-2025-A07. ZZ is thankful for support from the National Natural Science Foundation of China (grant no. 12233005).  MJV acknowledges a grant from the Finnish Society for Sciences and Letters. The authors also thank the reviewer for constructive comments.

\bibliography{ref}{}


\bibliographystyle{aasjournal}

\end{document}